\documentclass[aps,preprint,10pt,twocolumn]{revtex4}%
\usepackage{amsfonts}
\usepackage{amsmath}
\usepackage{amssymb}
\usepackage{graphicx}%
\begin{document}
\preprint{ }
\title{Magnetic Phase Transition in FeRh }
\author{R. Y. Gu and V. P. Antropov}
\affiliation{Condensed Matter Physics, Ames Laboratory, Ames, IA 50011}
\date{\today}

\begin{abstract}
Density functional calculations are performed to investigate the phase
transition in FeRh alloy. The effective exchange coupling, the critical
temperature of magnetic phase transition and the adiabatic spin wave spectrum
have been obtained. Different contributions to the free energy of different
phases are estimated. It has been found that the antiferro-ferromagnetic
transition in FeRh occurs mostly due to the spin wave excitations.

\end{abstract}
\maketitle

The antiferromagnetic (AFM) - ferromagnetic (FM) phase transition in FeRh with
the ordered CsCl structure has been intensively studied both experimentally
and theoretically. This transition occurs at T$_{\text{tr}}\approx340$ K
without any accompanying structural changes \cite{shirane}, although there is
an abrupt 1\% volume expansion. At low temperatures the magnetic configuration
of FeRh is the type-II AFM (successive layers of (111) Fe planes AFM coupled)
with moments $3.3\mu_{B}$ on Fe atoms \ (Rh atoms are nonmagnetic). Above
$T_{\text{tr}}$ in FM phase magnetic moments are $3.2\mu_{B}$ on Fe atoms,
$1.0\mu_{B}$ on Rh atoms and the Curie temperature ($T_{\text{C}}$) is
$\approx670$ K \cite{kouvel1}. In addition, it was found that $T_{\text{tr}}$
is increased with pressure \cite{wayne,annaorazov2}. The AFM-FM transition can
also be induced by applying external magnetic field, whose critical value at
zero temperature is about 300 kOe and smaller at higher temperatures
\cite{ponomarev}, making this material a natural magnetic multilayer with a
large magnetoresistance effect.

Early theories of this transition based on the exchange-inversion model
\cite{kittel}, which assumes a change of sign for the exchange parameter at
some volume, can not account for the experimental observation of the large
entropy changes at $T_{\text{tr}}$
\cite{ponomarev,kouvel,richardson,annaorazov}. After discovering that the
electronic specific heat in the FM phase is nearly four times larger than in
the AFM phase, Tu et al. \cite{tu} proposed that the change of the band
electron entropy plays a major role in this transition. However, they used
iron-rich alloys where the value of the specific heat is very sensitive to the
concentration \cite{ivarsson}. Consecutive dielectric function measurements
\cite{chen} demonstrated that the band structure of FeRh is not drastically
modified during the AFM-FM phase transition.

First-principles band-structure calculations were also carried out to study
this transition. Earlier calculations \cite{kulicov} did not compare the
relative stability of the AFM and FM states. Moruzzi and Marcus \cite{moruzzi}
confirmed that the type-II AFM structure is the ground state, while the FM
structure represents another stable solution with the total energy nearly
$\Delta E$=2 mRy/atom higher at a larger volume. Similar results were obtained
in Refs.\cite{szajek,gruner}. However, the energy difference $\Delta E$
obtained in these calculations appears to be much larger than the experimental
data and, overall these studies did not provide any convincing explanation of
the nature of this transition. The experimental $\Delta E$, deduced either
from the latent heat $T_{\text{tr}}\Delta S$ or from the critical magnetic
field at zero temperature \cite{ponomarev}, is about 0.2 mRy/atom, which is an
order of magnitude smaller than the calculated value. On the other hand, while
in Ref.\cite{moruzzi} the calculated equilibrium lattice constant of the FM
state is only 0.5\% larger than in AFM one, the energy of latter remains lower
than the energy of the FM state until the lattice constant is increased by
3\%. The authors\cite{moruzzi} proposed that the zero-point lattice vibrations
can correct the total energy result. However, from the Debye temperature
$\Theta_{D}$ calculated in that paper, one can find that the correction due to
zero-temperature vibration energy $E_{0}=9k_{B}\Theta_{D}/8$ is nearly two
orders of magnitude smaller than the calculated $\Delta E$. Gruner et al.
\cite{gruner} investigated the thermodynamic behavior of the system within
Ising model and found that at $T_{\text{tr}}$ the free energy, gained due to
thermoexcitation in the FM state, is nearly 0.02 mRy/atom larger than that in
the AFM state. So, they proposed that it is this thermoexcitation that drives
the transition. However, the magnitude of their free energy change appears too
small to compensate the internal energy loss for the transition to occur. So
far, in spite of many years of research, there has been no convincing
explanation of the nature of the phase transition in FeRh.

In this paper we study this transition using first principles calculations and
the non-collinear version of the linear muffin-tin orbital method in the
atomic-sphere approximation (LMTO-ASA). In the local spin density
approximation (LSDA) we use the von Barth-Hedin potential, and for the
nonlocal corrections the Langreth-Mehl-Hu functional\cite{langreth}, the
relativistic effects and the combined corrections \cite{skriver} are included.
For the radii of atoms, the ratio $R_{\text{Rh}}/R_{\text{Fe}}=1.03$ was used.
The self-consistent calculations are performed for the different lattice
parameters for FM and AFM states using a spin spiral approach. The gradient
corrections are expected to be important in Fe-rich BCC based systems due to
well known fact that LSDA predicts, for instance, the wrong FCC ground state
pure Fe\cite{bagno}.\ Below we will show that it is exactly a case in FeRh.
Using electronic density of states (DOS) and exchange parameters, we calculate
the free energy change of the AFM and FM states. Various thermal quantities
related to the AFM-FM transition, including the transition temperature, its
pressure dependence, the entropy and the specific heat changes are calculated
and compared with the corresponding experiments. According to our
calculations, the AFM-FM transition in FeRh appears primarily due to the
magnon (spin wave) excitations.

Fig. 1 (a) shows the calculated energy of the AFM and FM states. In LSDA the
equilibrium Wigner-Seitz radius $R_{\text{WS}}$ for the AFM (FM) phase is
$2.767$ $(2.780)$ a.u., which is smaller than the previous non-relativistic
result 2.782 (2.798) a.u.\cite{moruzzi} (our non-relativistic result is 2.789
(2.803) a.u.). The energy difference between the AFM and FM states at their
respective equilibrium $R_{\text{WS}}$ is $1.89$ mRy/atom. With nonlocal
corrections this energy difference is reduced to $0.206$ mRy/atom, which is in
agreement with the experimental value $0.196$ mRy/atom \cite{ponomarev}. Other
results are listed in Table I.

Only collinear AFM and FM configurations were observed in the experiments. In
our calculation, however, we found that without nonlocal corrections the FM
configuration is not a locally stable state with respect to the magnetic
moment deviations, as is shown in Fig. 1 (b) (two kinds of noncollinear
configurations are considered). In both those states the Fe atoms are divided
into two sublattices, identical to that in the type-I (successive (001) Fe
layers belong to the different sublattices) and type-II AFM states with the
angle $\theta$ between Fe magnetic moments from the different sublattices,
while the moments of Rh atoms are parallel to the sum of the Fe moments. Such
instability of the FM state is removed if the nonlocal corrections are taken
into account. Results from Fig. 1 (a) and (b) suggest that the nonlocal
corrections are important in FeRh and necessarily should be taken into account
in the investigation of the phase transition.

Using formalism from Ref.\cite{liechtenstein} we calculated the parameters of
the exchange coupling. The obtained parameters are different for the FM and
AFM configurations (see Table II). To estimate the magnetic contributions to
the free energy we calculated the magnon spectrum in both phases. In the
adiabatic approximation for the AFM state $\omega_{\mathbf{q}}^{\text{AFM}%
}=(2g\mu_{B}/m_{\text{Fe}})\sqrt{(J_{0}-J_{\mathbf{q}})(J_{0}-J_{\mathbf{q}%
+\mathbf{Q}})},$ with $J_{\mathbf{q}}=\sum_{j}J_{ij}e^{i\mathbf{q}%
\cdot\mathbf{R}_{ij}}$ being the Fourier transformation of $J_{ij}$ in AFM
state and $\boldsymbol{Q}=(\pi\pi\pi)$. For the FM state $\omega_{\mathbf{q}%
}^{\text{FM,}\pm}=g\mu_{B}(A_{\mathbf{q}}+B_{\mathbf{q}}\pm\sqrt
{(A_{\mathbf{q}}-B_{\mathbf{q}})^{2}+4X_{\mathbf{q}}^{2}})$, where
$A_{\mathbf{q}}=(J_{0}^{\text{FeFe}}-J_{\mathbf{q}}^{\text{FeFe}}%
+J_{0}^{\text{FeRh}})/m_{\text{Fe}}$, $B_{\mathbf{q}}=(J_{0}^{\text{RhRh}%
}-J_{\mathbf{q}}^{\text{RhRh}}+J_{0}^{\text{FeRh}})/m_{\text{Rh}}$, and
$X_{\mathbf{q}}=J_{\mathbf{q}}^{\text{FeRh}}/\sqrt{m_{\text{Fe}}m_{\text{Rh}}%
}$, with $J_{\mathbf{q}}^{\text{FeFe}}$, $J_{\mathbf{q}}^{\text{FeRh}}$, and
$J_{\mathbf{q}}^{\text{RhRh}}$ being the Fourier transformations of the
exchange interactions inside (or between) the corresponding sublattice(s). The
magnon DOS obtained from this spectrum is shown in Fig. 2 together with the
electronic DOS.

\ To study the relative stability of the AFM and FM configurations at finite
temperatures, let us compare their free energies. Here we consider the
contributions from electrons and magnons only. The lattice contribution is
neglected because the magnitudes of the bulk moduli and Debye temperatures in
both phases are very similar (Table I). The free energy due to band electrons
and magnons are given by
\begin{align}
F_{\text{el}}(T)  &  =\frac{1}{2}\{\varepsilon_{F}n-k_{B}T\int d\varepsilon
N(\varepsilon)\log(1+e^{(\varepsilon_{\text{F}}-\varepsilon)/k_{B}%
T})\}\ ,\nonumber\\
F_{\text{mag}}(T)  &  =-\frac{k_{B}T}{2}\int\frac{d\mathbf{q}}{(2\pi)^{3}}%
\log(1-e^{-\omega_{\mathbf{q}}/k_{B}T})\ ,
\end{align}
where $n$ and $N(\varepsilon)$ are the number of electrons and the electronic
DOS, correspondingly. Fig. 3 (a) shows the free energy difference $\Delta
F=F^{\text{FM}}-F^{\text{AFM}}$ as a function of temperature. The transition
temperature $T_{\text{tr}}$, determined from $\Delta F=0$, is $371$ K, which
is close to the experimental result $T_{\text{tr}}\approx340$ K. Both
contributions are shown, with the main one (more than 80\%) coming from
magnons. So, the origin of the AFM-FM transition should be attributed
primarily to the magnon excitations rather than to pure electronic spectrum
modifications. The obtained $\Delta F(T)$ also enables us to get the pressure
dependence of the transition temperature, whose experimental value is about
$dT_{\text{tr}}/dP\approx5.1\sim5.8$ K/kbar \cite{wayne,annaorazov2}. With
applied pressure the AFM state gains more free energy $\Delta G=P\Delta V$
($\Delta V$ is the volume difference) than the FM state and the derivative
$d\Delta G/dP$ is close to $-6.2\times10^{-3}$mRy/kbar per atom. From Fig. 3
(a) $d\Delta F/dT=-1.17\times10^{-3}$mRy/K at $T_{\text{tr}}$ and from the
equilibrium condition $d\Delta F=d\Delta G$ we obtain $dT_{\text{tr}%
}/dP\approx5.3$ K/kbar, which agrees well with the experiments.

From the obtained electronic DOS and magnon spectrum, one can also evaluate
various thermal quantities. In Fig.3 (b) we show the calculated differences of
the entropy and the specific heat between AFM and FM states as a function of
temperature. These two quantities are independent of the zero temperature
energy, with their measured values at $T_{\text{tr}}$ being $\Delta S^{\exp}$
$\approx13-19.6$ {Jkg}$^{-1}${K}$^{-1}$\cite{kouvel,richardson,annaorazov} and
$\Delta C^{\exp}\approx13.6$ {Jkg}$^{-1}${K}$^{-1}$ \cite{richardson}. Just as
the free energy, the calculated contributions to both $\Delta S$ and $\Delta
C$ near $T_{\text{tr}}$ mainly determined by the magnon excitations. At the
calculated $T_{\text{tr}}=371$ $K$, we obtain $\Delta S=19.3$ J{kg}$^{-1}$%
{K}$^{-1}$ and $\Delta C=15.1$ {Jkg}$^{-1}${K}$^{-1}$, while at $T_{\text{tr}%
}\approx340$ K, the corresponding values are $17.9$ and $15.6$ {Jkg}$^{-1}$%
{K}$^{-1}$.

Since the Rh atoms have a nonzero value ($1\mu_{B}$) of magnetic moments in
the FM state, this state has more magnetic degrees of freedom. It was proposed
that this additional number of degrees of freedom increases the entropy and
thus stabilizes the FM state\cite{moruzzi}. The evaluation of this entropy
gain gives $\Delta S\approx Nk_{B}\log2\approx36$ {Jkg}$^{-1}${K}$^{-1}$,
which is too large compared to the experimental results. According to our
calculated DOS of magnon, this picture of extra entropy is not quite accurate.
>From Fig. 2 (b) one can see that in the FM state the magnon excitations
$\omega_{\mathbf{q}}^{-}$ and $\omega_{\mathbf{q}}^{+}$ are separated by the
large energy gap. We call the $\omega_{\mathbf{q}}^{-}$ ($\omega_{\mathbf{q}%
}^{+}$) mode to be Fe(Rh)-like, because a similar mode (the dotted line in
Fig. 2(b)) can be obtained if we fix the orientations of the Rh (Fe) moments
by letting $X_{\mathbf{q}}=0$ in the expression for $\omega_{\mathbf{q}%
}^{\text{FM}}$. Near $T_{\text{tr}}$ only the Fe-like mode contributes to the
thermal properties so that the number of the effective magnon states in the FM
and AFM phases is the same, i.e., there is essentially the same number of spin
degrees of freedom in the FM and AFM states. That does not mean, however, that
the nonzero Rh moments in the FM phase do not contribute to the thermal
properties. On Fig. 2(b) it is shown that without the movement of the Rh
moments there is the energy gap between the ground state and the lowest-energy
Fe-like magnons $\omega_{\mathbf{q=0}}^{-}$ . A calculation shows that at
$T_{\text{tr}}$\ the magnon free energy (with the orientations of the Rh
moments being fixed) is only about one third of that when they are not fixed.
In other words, the Rh moments soften considerably the stiffness of the
Fe-like magnons, significantly influencing the thermal properties. A
comparison of the magnon DOS in the AFM and FM states indicates that near
$T_{\text{tr}}$ it is much easier to excite the Fe-like magnons in the FM
state. This is the main reason for the difference in thermal properties
between the two phases, and it is also the driving force of the AFM-FM
transition in FeRh.

Finally let us evaluate the Curie temperature by using the obtained pair
exchange interactions $J_{ij}$. In the mean field\ (MF) approximation
\begin{equation}
T_{C}^{\text{MF}}=\frac{1}{3k_{B}}(J_{0}^{\text{FeFe}}+J_{0}^{\text{RhRh}%
}+\sqrt{(J_{0}^{\text{FeFe}}-J_{0}^{\text{RhRh}})^{2}+4(J_{0}^{\text{FeRh}%
})^{2}}\ )\ .
\end{equation}
From $J_{0}^{\text{FeFe}}=-3.56$ mRy, $J_{0}^{\text{FeRh}}=9.92$ mRy and
$J_{0}^{\text{RhRh}}=0.85$ mRy we obtain 927K, which is nearly 40\% higher
than the experimental $T_{C}\sim670$K.\ Our Monte-Carlo calculations (with all
calculated long ranged $J_{ij}$ included) produced correspondingly $660-690$
K, so that the ratio $T_{C}/T_{C}^{\text{MF}}\approx0.71-0.74$ is nearly the
same as that of the simple cubic lattice Heisenberg model where $T_{C}%
/T_{C}^{\text{MF}}=0.722$\cite{peczak}. The agreement of the calculated and
the experimental $T_{C}$ indicates that the Heisenberg model may still work
well in the temperature region near $T_{C}$ in FeRh.

RYG thanks Dr. G. D. Samolyuk for his help with the computational codes. This
work was carried out at the Ames Laboratory, which is operated for the U.S.
Department of Energy by Iowa State University under Contract No. W-7405-82.
This work was supported by the Director for Energy Research, Office of Basic
Energy Sciences of the U.S. Department of Energy.

\bigskip\newpage

Fig.1. The total energy obtained in the local \ (L) and nonlocal (NL)
approximations, represented by open and solid symbols. The total energy for:
(a) AFM (circles) and FM (squares) states as a function of $R_{\text{WS}}$;
(b) for the type-I (squares) and type-II (circles) noncollinear states as a
function of the spin spiral angle $\theta$. $R_{\text{WS}}$ is fixed at the
equilibrium value of the FM state.

\ Fig.2. Calculated DOS \ of FM (solid line) and AFM (dashed line) states: (a)
the electronic DOS and (b) the magnon DOS. In (b) the dotted lines correspond
to the FM magnon DOS when $X_{\mathbf{q}}=0$ in the expression for
$\omega_{\mathbf{q}}^{\text{FM}}$.

Fig.3. The calculated differences of (a) the free energy $\Delta F,$ (b) the
entropy $\Delta S$ and the specific heat $\Delta C$ between FM and AFM phases.
The dashed and dotted lines (short-dashed and short-dotted lines for $\Delta
C$) correspond to the contributions from the electrons and magnons,
correspondingly. The solid line is their sum.

\newpage

\begin{table}[ptb]
\caption{Calculated physical properties of the AFM and FM configurations of
FeRh obtained in the local (first row) and nonlocal approximations.
$m_{\text{Fe}}$ and $m_{\text{Rh}}$ are magnetic moments of Fe and Rh atoms, B
is the bulk modulus, $\Theta_{D}$ is the Debye temperature and $N(\epsilon
_{F})$ is DOS per formula unit at the Fermi level.}%
\begin{tabular}
[c]{cccccccc}%
$R_{\mbox{\scriptsize WS}}$ & $\Delta E$ & $m_{\mbox{\scriptsize Fe}}$ &
$m_{\mbox{\scriptsize Rh}}$ & B & $\Theta_{D}$ & $N(\epsilon_{F})$ & \\
(a.u.) & (mRy/atom) & ($\mu_{B}$) & ($\mu_{B}$) & (kbar) & (K) & (States/Ry) &
\\\hline
2.767(2.780) & 1.89 & 3.12(3.22) & 0(1.04) & 2454(2364) & 385(379) &
18.0(29.5) & \\\hline
2.796(2.807) & 0.206 & 3.28(3.31) & 0(1.02) & 2194(2181) & 366(365) &
15.6(28.0) &
\end{tabular}
\end{table}

\begin{table}[ptb]
\caption{Pair exchange parameters (in mRy) in AFM and FM phases of FeRh.
Corresponding coordinates are shown in units of the lattice constant.}%
\begin{tabular}
[c]{cccccc}%
Type of pair & $\Delta x$ & $\Delta y$ & $\Delta z$ & $J_{ij}%
^{\mbox{\scriptsize FM}} $ & $J_{ij}^{\mbox{\scriptsize AF}} $\\\hline
Fe-Rh & 0.5 & 0.5 & 0.5 & 1.062 & 0\\
\  & 1.5 & 0.5 & 0.5 & 0.058 & 0\\\hline
\  & 1 & 0 & 0 & -0.098 & 0.442\\
Fe-Fe & 1 & 1 & 0 & 0.104 & 0.008\\
\  & 1 & 1 & 1 & -0.479 & 0.603\\
\  & 2 & 0 & 0 & 0.120 & 0.099\\
\  & 2 & 1 & 0 & 0.045 & 0.005\\\hline
Rh-Rh & 1 & 0 & 0 & 0.086 & 0\\
\  & 1 & 1 & 0 & 0.018 & 0
\end{tabular}
\end{table}


\begin{thebibliography}{99}                                                                                               %


\bibitem {shirane}G. Shirane, R. Nathans, and C. W. Chen, Phys. Rev.
\textbf{134}, A1547 (1964).

\bibitem {kouvel1}J. S. Kouvel, and C. C. Hartelius, J. Appl. Phys. Suppl.
\textbf{33}, 1343 (1962).

\bibitem {wayne}R. C. Wayne, Phys. Rev. \textbf{170}, 523 (1968).

\bibitem {annaorazov2}M. P. Annaorazov, J. Alloys Comp. \textbf{354}, 1 (2003).

\bibitem {ponomarev}B. K. Ponomarev, Zh. Eksp. Teor. Fiz. \textbf{63}, 199
(1972) [Sov. Phys.-JETP\textbf{36}, 105 (1973)].

\bibitem {kittel}C. Kittel, Phys. Rev. \textbf{120}, 335 (1960).

\bibitem {kouvel}J. S. Kouvel, J. Appl. Phys. \textbf{37}, 1257 (1966).

\bibitem {richardson}M. J. Richardson, D. Melville, and J. A. Ricodeau, Phys.
Lett. \textbf{46A}, 153 (1973).


\bibitem {annaorazov}M. P. Annaorazov, S. A. Nikitin, A. L. Tyurin, K. A.
Asatryan, and A. K. Dovletov, J. Appl. Phys. \textbf{79}, 1689 (1996).


\bibitem {tu}P. Tu, A. J. Heeger, J. S. Kouvel, and J. B. Comly, J. Appl.
Phys. \textbf{40}, 1368 (1969).

\bibitem {ivarsson}J. Ivarsson, G. R. Pickett, and J. Toth, Phys. Lett.
\textbf{35A}, 167 (1971).

\bibitem {chen}L. Y. Chen and D. W. Lynch, Phys. Rev. B \textbf{37}, 10503 (1988).

\bibitem {kulicov}N. I. Kulikov, E. T. Kulatov, L. I. Vinokurva, and M.
Pardavi-Horvath, J. Phys. F \textbf{12}, L91 (1982); C. Koenig, \textit{ibid.}
\textbf{12}, 1123 (1982).

\bibitem {moruzzi}V. L. Moruzzi and P. M. Marcus, Phys. Rev. B \textbf{46},
2864 (1992).

\bibitem {szajek}A. Szajek and J. A. Morkowski, Physica B \textbf{193}, 81 (1994).

\bibitem {gruner}M. E. Gruner, E. Hoffmann, and P. Entel, Phys. Rev. B
\textbf{67}, 064415 (2003).

\bibitem {bagno}P. Bagno, O. Jepsen, and O. Gunnarsson, Phys. Rev. B
\textbf{40}, 1997 (1989).

\bibitem {moriya}T. Moriya, \textit{Spin Fluctuations in Itinerant Electron
Magnetism} (Springer, Berlin, 1985).

\bibitem {skriver}H. L. Skriver, \textit{The LMTO method} (Spring-Verlag,
Berlin, 1984).

\bibitem {langreth}D. C. Langreth and M. J. Mehl, Phys. Rev. Lett.
\textbf{47}, 446 (1981).

\bibitem {liechtenstein}A. I. Liechtenstein, M. I. Katsnelson, V. P. Antropov,
and V. A. Gubanov, J. Magn. Magn. Mater \textbf{67}, 65 (1987); V. P.
Antropov, B. N. Harmon and A. N. Smirnov, \textit{ibid.} \textbf{200}, 148 (1999).

\bibitem {halilov}S. V. Halilov, H. Eschrig, A. Y. Perlov, and P. M. Oppeneer,
Phys. Rev. B \textbf{58}, 293 (1998).

\bibitem {peczak}P. Peczak, A. M. Ferrenberg, and D. P. Landau, Phys. Rev. B
\textbf{43}, 6087 (1991).
\end{thebibliography}
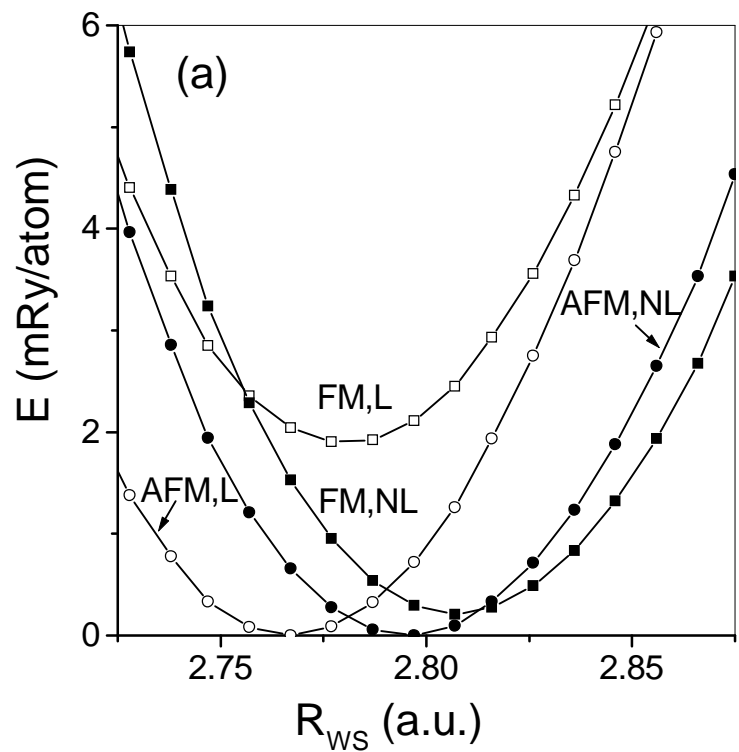
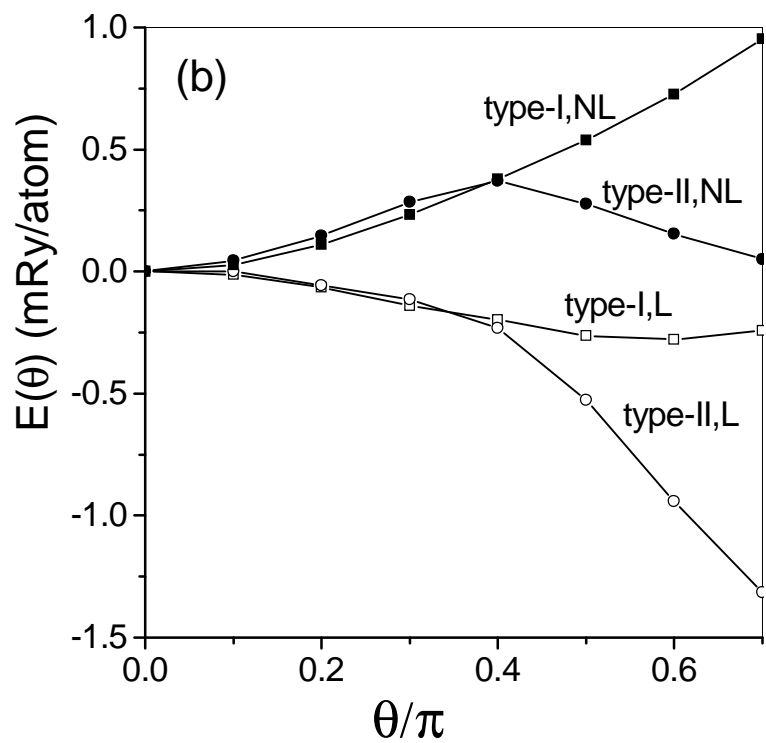
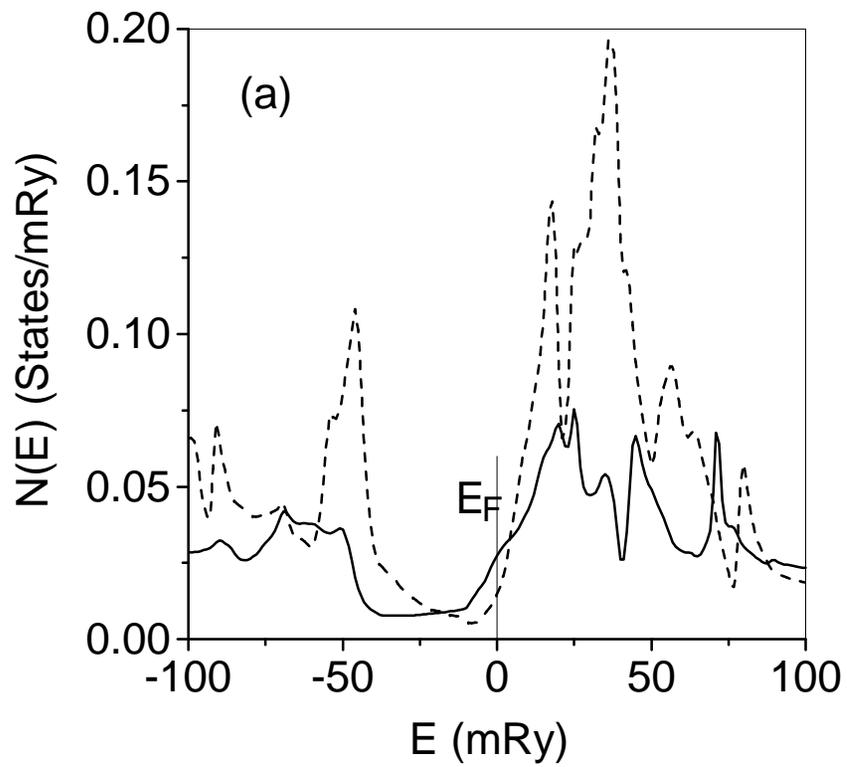
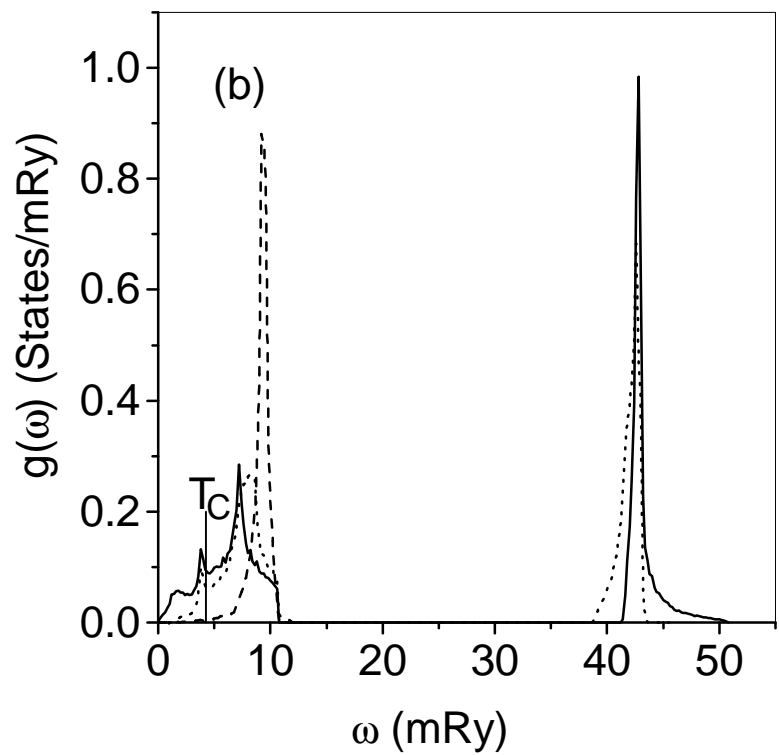
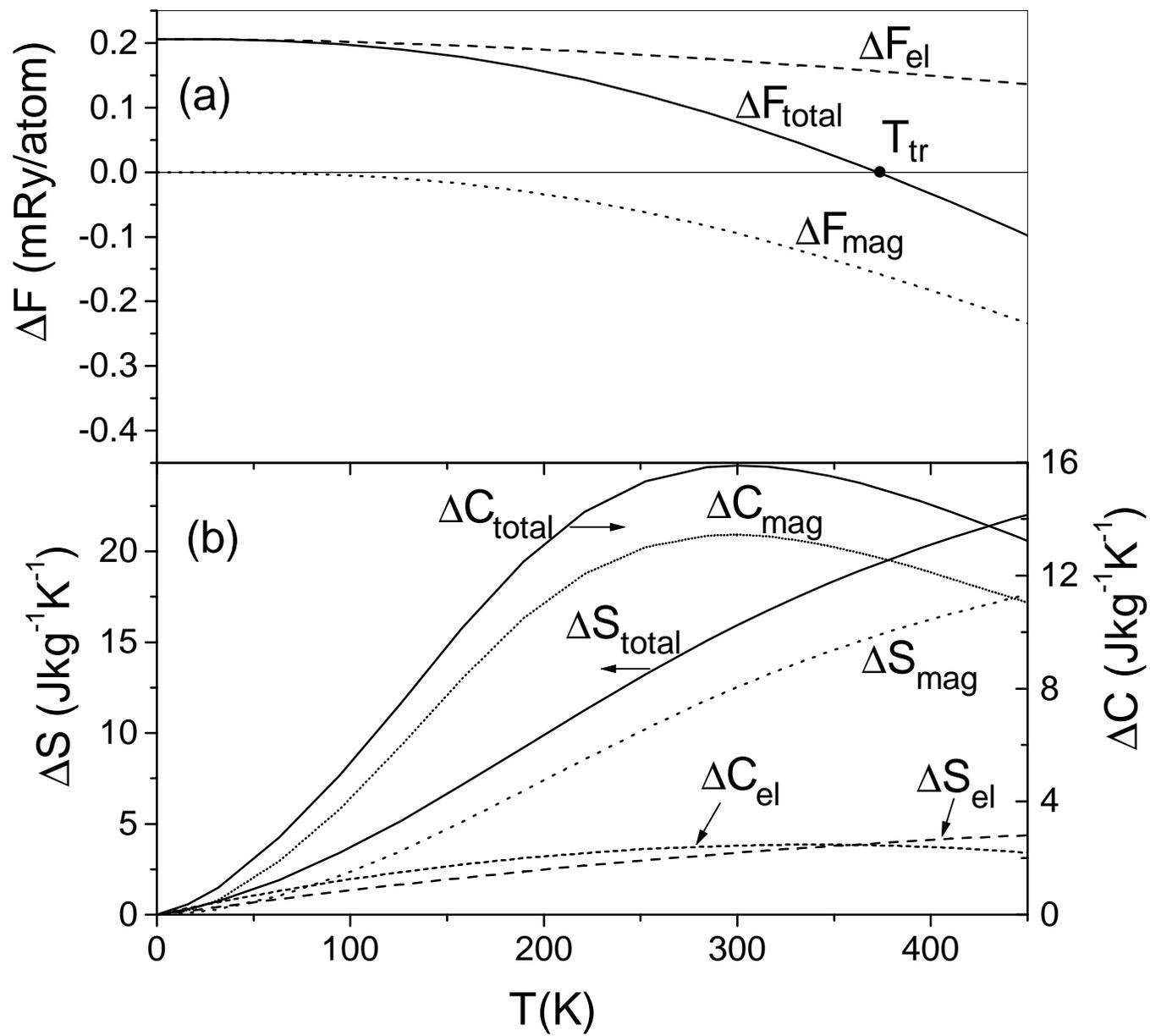
\end{document}